\documentclass[11pt,a4paper]{article}
\usepackage{eepic}
\usepackage[]{epic,eepic,epsf}
\usepackage[xdvi]{graphicx}

\newcommand{\be}{\begin{equation}} 
\newcommand{\ee} {\end{equation}} 
\newcommand{\ob} {\overline} 
 
\begin{document}  
 
\flushbottom

\title{Continuously varying  exponents in a   
sandpile model with dissipation near surface}

\author{S. L\"ubeck$^1$ and D. Dhar$^2$\\ 
\footnotesize $^1$ Theoretische Tieftemperaturphysik,  
Gerhard-Mercator-Universit\"at Duisburg,\\  
\footnotesize Lotharstr. 1, 47048 Duisburg, Germany \\ 
\footnotesize $^2$ Department of Theoretical Physics,  
Tata Institute of Fundamental Research,\\ 
\footnotesize Homi Bhabha Road, MUMBAI, 400005, India\\}

\date\today 
 
\maketitle 
 
\begin{abstract} 
 
We consider the directed Abelian sandpile model in the presence of sink 
sites whose density $f_t$ at depth $t$ below the top surface varies as 
$c\,  t^{-\chi}$. For $\chi>1$ the disorder is 
irrelevant. For $\chi <1 $, it is relevant and the model is no longer 
critical for any nonzero $c$. For $\chi=1$ the 
exponents of the avalanche distributions depend continuously on the 
amplitude $c$ of the disorder. We calculate this 
dependence exactly, and verify the results with simulations. 


\vspace{-15cm}
\hspace{-1.6cm}
\begin{minipage}{14.5cm}
TIFR preprint no.  TIFR/TH/ 00-34
\hfill 
accepted for publication in J.\,Stat.\,Phys.
\end{minipage} 
\vspace{14.1cm}
 
\end{abstract}


\section{Introduction} 
 
An important question in the general area of self-organized criticality 
\cite{bak} is the universality of critical exponents. 
Do small perturbations of the local evolution rules 
leave the critical exponents unchanged? 
How do perturbations of the system at the boundary 
affect  the critical steady state? These are the questions we study 
in this paper, and construct a simple analytically tractable model where 
changes near the boundary lead to {\it continuously varying critical 
exponents}. 
  
The influence of boundary perturbation in nonequilibrium systems is more 
complicated than in equilibrium systems \cite{derrida}. In the latter, 
away from phase transition points, the bulk is largely unaffected by the 
boundary, except that in case of spontaneous symmetry breaking, the 
boundary conditions will remove the degeneracy between different phases, 
and some finite-size effects. {\it Near critical phase transitions}, the 
critical behavior near the boundary is substantially modified, and 
constitutes a different class of (surface) critical phenomena, 
characterized by a new set of (surface) exponents \cite{igloi}.  In the 
so-called `ordinary transitions', the surface phase-transition is driven 
by the bulk transition in the sense that the transition occurs at the 
critical temperature corresponding to the bulk, and its position or the 
critical exponents are not affected by local changes in the Hamiltonian 
near the surface.

In self-organized critical systems, the correlation length is, by 
definition, always large, and the influence of perturbations at the 
boundary is felt deep inside. In the well-known Bak-Tang-Wiesenfeld (BTW) 
sandpile model of self-organized criticality~\cite{bak87}, 
even the {\it existence} of 
a critical steady state depends on the presence of a surface where 
particles can leave the system.  Thus, in such systems, boundary 
perturbations affect the overall behavior of the system to a much greater 
extent.  As the steady state is critical, various measured average 
quantities in the steady state near a surface will differ from the bulk 
values by an amount which {\it will decrease as a power law of the 
distance from the surface}. 
 
For example, in the BTW model in two dimensions, Brankov {\it et 
al.}~\cite{BIP93} have shown that the mean height of the pile at a site at 
distance  $t$ from the surface is less than the mean value in the bulk by 
an amount proportional to $t^{-2}$.  In a one dimensional sandpile model with 
threshold dissipation, Ali found that the density is larger near the 
boundary \cite{ali} and the excess density varies as $1/t$. We would like 
to understand the effect of such extended perturbations on other 
properties of the system. For example, one may ask if such a density 
profile significantly affects the probability of extinction of avalanches 
that reach the boundary. One could even ask if it {\it determines} the 
avalanche exponents \cite{stella}. In this context, it seems worthwhile to 
understand the properties of self-organized critical systems with 
`artificially produced' power-law surface perturbations.

In this paper, we consider the directed Abelian sandpile model in the 
presence of some sink sites, whose density varies with the distance from 
the surface with a power $\chi$. We are able to determine the 
$\chi$-dependence of the avalanche exponents exactly. Our analysis shows 
that the disorder is irrelevant if $\chi >1$, and is relevant for $\chi < 
1$. For $\chi = 1$, the perturbation is marginal, and the critical 
exponents depend continuously on the disorder strength.  In the case of 
equilibrium critical phenomena, for Ising model with deviation of the 
coupling strength from the critical value varying with distance from 
surface, a similar behavior is seen \cite{hilhorst}, and may be understood 
in terms of a simple renormalization argument \cite{burkh82}.

In fact precisely this question comes up in the sandpile model studied by 
Tadi{\'{c}} and Dhar (TD) \cite{tadic}.  This model (its definition is 
given later in the paper) belongs to the universality class of {\it 
directed models with stochastic toppling rule, with no multiple topplings} 
\cite{vesp}. For this model, TD presented some intuitive arguments to 
determine the exact critical exponents for a general dimension $d$ in 
terms of the exponents of the directed percolation problem. The arguments, 
involving determining the statistics of clusters in a directed percolation 
model with space-varying density, are plausible, but not rigorous. In the 
model studied here, our exact answer coincides with that obtained based on 
the same reasoning, thus provides some additional evidence in the latter's  
favor.

In the next section we define our model.  We determine the 
disorder-averaged two-point correlation function of toppling events 
exactly in section~\ref{sec:two_point}.   
This shows the qualitative difference in the 
behaviors of the avalanche distribution for $\chi >1$ and $\chi < 1$. We 
give heuristic arguments to determine the avalanche exponents as a 
function of the amplitude $c$ in the marginal case 
$\chi =1$. These results are confirmed by numerical investigations.  A 
somewhat more careful derivation of the avalanche exponents requires the 
consideration of the three-point correlation function of toppling events 
which is presented in section~\ref{sec:three_point}. The connection to  
the argument  used in \cite{tadic}, and some generalizations 
to other systems are discussed in the concluding 
section.

\section{Definition of the  model} 
\label{asm_model} 
 
\begin{figure}[t] 
\begin{center} 
\includegraphics[width=8cm,angle=0]{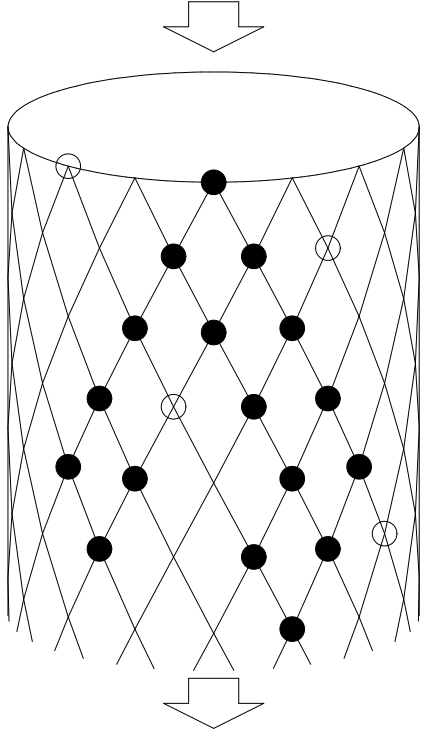}
\caption{The lattice structure of the directed Abelian 
sandpile model. Sand-grains are added at the top and removed 
from the bottom. Periodic boundary conditions are used in the 
horizontal direction. The full circles correspond to toppled lattice 
sites and the open circles mark sinks where sand-grains leave the 
system in the bulk. 
\label{fig:geo_01}} 
\end{center} 
\end{figure}

We consider a two dimensional lattice in the shape of an open cylinder 
(see Fig.~\ref{fig:geo_01}). Each site is labelled by two integer 
coordinates $(x,t)$, where $x$ labels the coordinate in the horizontal 
direction, and $t$ denotes the depth of the layer ($0 \leq t \leq T$). It 
is convenient to work with only the set of sites for which $ x+t $ is 
even (one of the two sublattices into which this lattice may be 
decomposed). We denote the number of sites in each horizontal row by $L$. 
To each lattice site $(x,t)$, a non-negative integer variable $h(x,t)$ is 
associated, called the height of the sandpile at that site.  A sand-grain 
is added at a randomly chosen site $x$ on the top layer ($t=0$).  If the 
height of a site $(x,t)$ exceeds the critical value $h_{\rm c}=2$, i.e, if 
$h(x,t)\ge 2$, it is unstable and it topples.  On toppling at a site, its 
height is decreased by $2$, and one grain is transferred to each of the 
two downward neighbors $(x \pm 1,t+1)$. In this way the transferred 
sand-grains may activate the two downward sites and an avalanche of 
relaxation events may take place. Some  typical avalanches are shown in 
Fig.~\ref{fig:aval_01}. 
 
The size of  an avalanche may be measured in terms 
by its downward  length (called its duration~$t$), or the number  
of toppled sites (called its mass $m$). In the critical steady state, 
the probability distribution of 
avalanche duration and avalanche mass  
has power-law tails~\cite{DR89}, i.e., 
 
\begin{equation} 
{\rm Prob}\{ duration \geq t\}  \sim \; t^{-\tau_t + 1} \quad \quad {\rm 
with}  
\quad \quad \tau_t=\frac{3}{2}, 
\label{eq:prob_t_exact} 
\end{equation} 
\begin{equation} 
{\rm Prob}\{ mass \geq m\} \sim \; 
m^{-\tau_m +1} \quad \quad {\rm with}  
\quad \quad \tau_m=\frac{4}{3}. 
\label{eq:prob_a_exact} 
\end{equation}

In this paper, we consider the effect of introducing surface disorder in 
the model. We do this by declaring a small fraction of sites as sink 
sites. Any particle falling onto a sink site is lost from the system. The 
sink sites are uncorrelated with each other. We assume that the 
concentration of sink sites $f_t$ at the depth~$t$ decreases as  
a power of~$t$: 
 
\begin{equation} 
f_t \;  = \; c \;\; t^{-\chi},  
\quad {\rm for} \quad t \gg 1. 
\label{eq:density_holes_l} 
\end{equation}

The case where the sink concentration is uniform was studied earlier by 
Tadi{\'{c}} {\it et al.}~\cite{TNURP92}. They showed numerically that any 
finite concentration of sinks destroys the criticality of the steady 
state.  Theiler~\cite{THEILER_1} showed that in the mean-field theory a 
uniform sink density $f$ induces an exponential cutoff in the depth 
reached by avalanches which varies as $f^{-1}$ for small $f$.

In the next sections we investigate the two-point and  
three-point correlation functions which allow to derive the  
avalanche exponents exactly.

\section{The two-point correlation function} 
\label{sec:two_point} 
 
The characterization of the steady state is very simple in our model. From 
the general theory of Abelian sandpile models \cite{asm}, it is easy to 
see that all stable configurations are recurrent, and occur with equal 
probability in the steady state. Thus, in the steady state, each non-sink 
site is empty or occupied with equal probability, independent of the 
occupation of other sites. 
 
\noindent Let us now consider the propagation of avalanches in the system. Let 
$G(x,t| x',t')$ be the probability in the steady state that a toppling 
will occur at $(x,t)$ if a particle is added at $(x',t')$. If the site 
$(x,t)$ is a sink site, we define $G(x,t|x',t')$ to be zero.  In the 
steady state, at non-sink sites, the average number of particles coming 
into $(x,t)$ must equal the number of leaving particles.  Thus, the 
two-point correlation function $G(x,t|x't')$ satisfies the equation 
\be  
G(x,t+1|x',t') = \frac{1}{2} \, \eta(x,t+1) \; \left [ G(x-1,t|x',t') + 
G(x+1,t|x',t'){\vphantom{\ob G}} \right ],  
\quad {\rm for} \quad t \ge t', 
\ee  
where $\eta(x,t+1)$ is zero if the site is a sink 
site, and one otherwise. Averaging over the configuration of sink sites 
gives the disorder-averaged two-point correlation function 
$\ob{G}(x,t+1|x',t')$.  Using $\langle \eta(x,t+1) \rangle = (1-f_{t+1})$  
and the 
fact that $\eta(x,t+1)$ is independent of $G(x \pm 1,t|x',t')$, it is easy 
to see that $\ob{G}(x,t+1|x',t')$ satisfies exactly the difference equation 
\be  
\ob{G}(x,t+1|x',t') \; = \;\frac{1}{2} \, ( 1- f_{t+1}) \;   
\left [ \ob{G}(x-1,t|x',t') + \ob{G}(x+1,t|x',t') \right ], 
\quad {\rm for} \quad t \ge t'.  
\label{eq:recurs_01}  
\ee 
The disorder-averaged correlation function $\ob{G}$ is 
translationally invariant in the $x$-direction, i.e.,  
\be 
\ob{G}(x+x_{\scriptscriptstyle 0},t|x'+x_{\scriptscriptstyle 0},t') 
\; = \, \ob{G}(x,t | x', t') ,
\ee  
for all $x_{\scriptscriptstyle 0}$.
We use 
Eq.(\ref{eq:recurs_01}) as a recursion equation and determine 
$\ob{G}(x,t| 0,0)$ for all $t$, given its initial value at $t=0$  
\begin{equation}  
\ob{G}(x,0 | 0,0) = \frac{1}{2} \, ( 1 - f_{0})\, \delta (x).   
\end{equation}  
It is easy to show that the disorder-average function $\ob{G}$ is given by 
\be \ob{G}(x,t| 0,0) = {\cal{P}}_{t}\; \; G_0(x,t|0,0) 
\label{eq:G_disorder}  
\ee  
where $G_0$ is the propagator for the problem with no sinks, i.e.,
\be  
G_0(x, t | 0, 0) \; = \; 2^{-t-1} \, 
\left ( 
t \atop \frac{x+t}{2}
\right )
\ee  
and where ${\cal{P}}_{t}$ is defined to be the product 
\be  
{\cal{P}}_{t} = \prod_{j=0}^{t} \, (1- f_j). 
\ee  
 
For the case with no disorder, it is known that  
${\rm Prob}\{ duration 
\geq t \}$ and  
$G_0(0,t|0,0)$ both vary with the same power of $t$~\cite{DR89}.   
If we make the plausible assumption that this continues to hold in the 
presence of sinks, we expect that the probability distribution  
of the avalanche duration is of the form 
\be  
{\rm Prob}\{ duration \geq t \} \; \sim \; t^{-1/2} \; {\cal P}_{t}\, 
\label{eq:prob_t_exact_dis}  
\ee  
since the propagator $G_0$ is known to decrease asymptotically  
as $t^{-1/2}$ 
for large $t$~\cite{DR89}. 
 
Let us consider some limiting cases.  In the case of homogeneous disorder 
$f_t=f$ we get that ${\cal{P}}_{t} \sim (1- f)^{t}$, and the probability 
that an avalanche reaches depth $t$ decreases exponentially with $t$. 
This is in agreement with the numerical results of \cite{TNURP92}. 
  
\begin{figure} 
\begin{center} 
\includegraphics[width=10cm,angle=0]{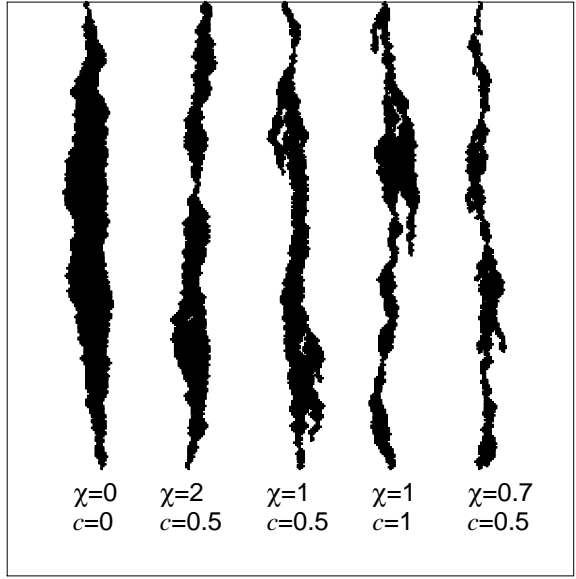}
\caption{Snapshot of five avalanches for various values of $\chi$ and 
$c$. In all cases the avalanche duration 
is $t=500$. In order to display the details the avalanches are stretched 
by the factor 2  in the horizontal direction. 
\label{fig:aval_01}} 
\end{center} 
\end{figure}

In the case that the disorder density $f_t$ decreases as a power-law 
according to Eq.~(\ref{eq:density_holes_l})  one has to distinguish three 
cases: 
 
$i)$ For $\chi > 1$, the disorder is irrelevant.  ${\cal{P}}_{t}$ tends to 
a finite constant value for $t\to \infty$, and the asymptotic behavior of 
the disorder averaged correlation function $\ob{G}$ is the same as that of 
$G_0$, i.e., the disorder does not affect the scaling behavior. A typical 
avalanche for $\chi=2$ is shown in Fig.~\ref{fig:aval_01}. While 
branchings can occur, most of the longer avalanches have only one 
prominent branch.

$ii)$ For $\chi < 1$, the disorder is relevant, and leads to loss of 
criticality in the system.  ${\cal{P}}_{t}$ tends to zero as  
$\exp{(-K 
t^{1-\chi})}$, where $K$ is a constant which depends on $\chi$ and 
$c$.  Thus the function $\ob{G}$ decreases to zero 
with the distance as a stretched exponential.  
In contrast to the case $\chi>1$ the avalanche 
propagation for $\chi<1$ is characterized 
by many branchings, though only one branch survives for long (see 
Fig.~\ref{fig:aval_01}).

\begin{figure} 
\begin{center} 
\includegraphics[width=10cm,angle=0]{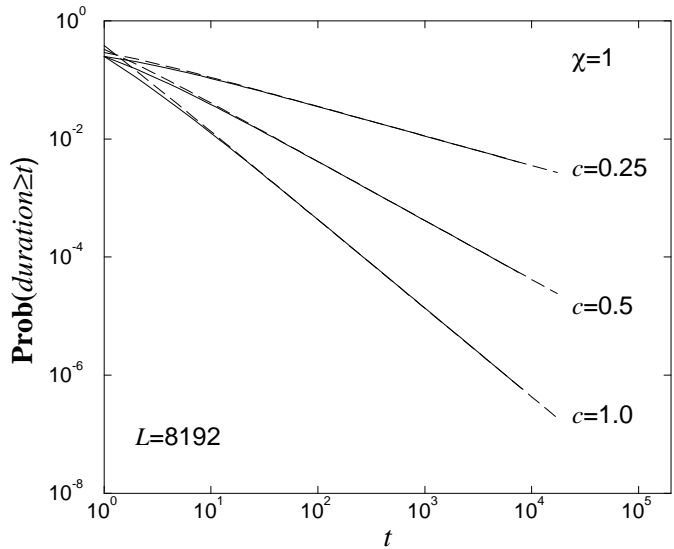}
 \caption{The probability distribution ${\rm Prob}\{ duration \geq t\}$ for $\chi=1$ 
          and various values of $c$ 
          for a system of size $L\times L$. 
          The dashed lines correspond to the derived scaling 
          behavior [Eq.~(\protect\ref{eq:prob_t_exact_dis})]. 
          The probability distributions ${\rm Prob}\{ duration \geq t\}$ are  
          averaged over $50$ different representations of the disorder    
          (hole-configuration) and $2\; 10^7$ non-zero-avalanches 
          for each disorder representation. 
 \label{fig:prob_t_xe1}} 
\end{center} 
\end{figure}

$iii)$ For $\chi =1$, the disorder is marginal and the 
avalanche distribution exponent varies continuously with the  
parameter $c$. 
Here, ${\cal{P}}_{t}$ tends to zero for large~$t$ as $t^{-c}$.  
From Eq.~(\ref{eq:G_disorder}), we see that 
$\ob{G}$ decreases as $t^{-1/2-c}$ and thus we find 
\begin{equation} \tau_t \; = \;  \frac{3}{2} \, + \, c . 
\label{eq:prob_t_dis_xe1_01}  
\end{equation}  
The avalanche distribution ${\rm Prob}\{ duration \ge t\}$ still exhibits 
a power-law 
decay, but the exponent depends continuously on the amplitude 
$c$. This behavior has been verified in our 
simulations.  Figure~\ref{fig:prob_t_xe1} displays the probability 
distribution ${\rm Prob}\{ duration \ge t\}$ for three different values of 
$c$. For sufficiently large durations the 
asymptotic behavior agrees with Eq.~(\ref{eq:prob_t_dis_xe1_01}).

Let us now consider the distribution of masses of avalanches. We see from 
the explicit form of the avalanche propagator $\ob{G}$ that while its 
dependence on $t$ is modified by the disorder, its dependence on the 
transverse coordinate $x$ is unaffected. Hence, even in the presence of 
disorder, the average transverse size of an avalanche at depth $t$ scales 
as $t^{1/2}$.  As each site in an avalanche topples only once, the average 
width of an avalanche cluster varies as $t^{1/2}$ and the total number of 
toppled sites $m$ scales as $t^{3/2}$.   
Then ${\rm Prob}\{ mass \geq m\} \; \sim \; m^{-\tau_m +1}$ with 
\be 
\tau_m\; = \; \frac{4}{3} \, + \, \frac{2}{3} \, c. 
\label{eq:tau_m_dis} 
\ee

\begin{figure} 
\begin{center} 
\includegraphics[width=10cm,angle=0]{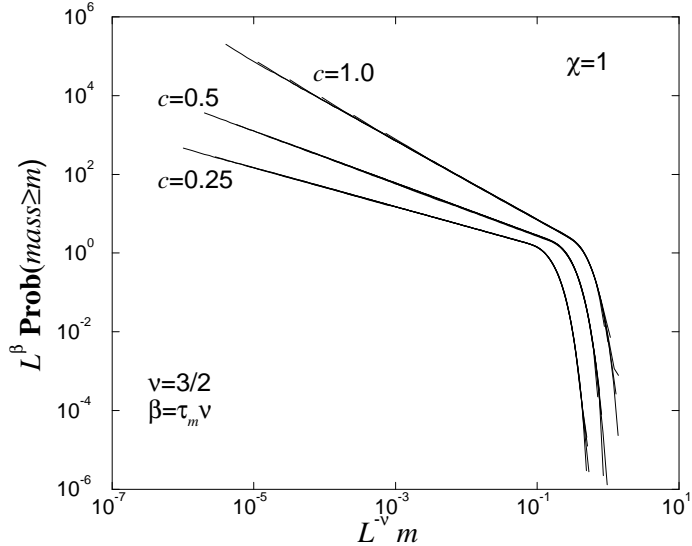}
\caption{Finite-size scaling analysis of the mass probability distribution 
${\rm Prob} \{ mass \ge m\}$ for $\chi=1$, three different values of 
$c$ and for various system sizes $L\in 
\{256,512,...,8192\}$. To avoid an overlap the curves for 
$c=0.5$ and $ c =0.25$ are 
shifted in the lower-left direction. 
 \label{fig:prob_m_xe1}} 
\end{center} 
\end{figure}

In our simulations, we used lattices of size $L\times T$ with $T=L$.
In this case, the transverse size is effectively infinite as even 
the largest avalanches are not affected by the finite transverse size. 
Thus the finite-size effects are governed by the longitudinal size ($T$) 
alone, and the mass distribution is expected to fulfill the finite-size 
scaling ansatz~\cite{KADANOFF_1}  
\begin{equation} 
{\rm Prob}(mass \geq m|T)  \sim \; T^{-\beta} \,  f(T^{-\nu} m). 
\end{equation} 
If finite-size scaling works a data collapse of the distributions for 
different system sizes~$L\times T$ (with $L=T$) 
is obtained for $\nu=3/2$ and $\beta = 1/2 + c$.  
The corresponding scaling plots for three 
different values of $c$ are shown in 
Fig.~\ref{fig:prob_m_xe1}. In all cases we observed good data collapse 
which confirms the above analysis.

\section{The three-point correlation function} 
\label{sec:three_point} 
 
In the previous section we obtained the avalanche exponents by using some 
scaling arguments, which are not careful in distinguishing the probability 
that a given site at depth $t$ is toppled in an avalanche from the 
probability that {\it at least} one site at depth $t$ is toppled.  We now 
present a cleaner derivation of the above results. 
 
Let us denote the number of sites toppling in the layer $t$ by $N(t)$.  
Based on the experience with the model without disorder, we make the 
assumption that the distribution of avalanches in this model can be 
described by the scaling function 
\be 
P(N(t)=n) \; \sim   \; A \, t^{-a} \, {\cal F}  (n/t^{b}) 
\label{eq:scal_assump_01} 
\ee 
for $n \neq 0$ and where the scaling function ${\cal F}(x)$  
decreases fast for large~$x$, and $A$ is a constant.  From 
this scaling form, it follows that  
$\langle N(t) \rangle \sim t^{2 b - a}$,  
and $\langle N^2(t) \rangle \sim t^{3 b - a}$.  
Thus knowing the $t$-dependence of $\langle N(t) \rangle$ and 
$\langle N^2(t) \rangle$, we can determine the exponents $a$ and $b$. 
Clearly, using Eq.(\ref{eq:G_disorder}) we get 
\be 
\langle N(t) \rangle  \; = \;  \sum_x \; \ob{G}(x,t|0,0) =  
\; \frac{1}{2} \; {\cal{P}}_{t}. 
\ee 
We would now like to calculate $\langle N^2(t) \rangle$.  Let 
$\ob{G}^{(3)}(x_1,x_2,t)$ be the probability that both sites $x_1$ and 
$x_2$ at depth $t$ topple when a particle is added at $x=0$ on the top 
layer $t=0$, averaged over all configurations of the disorder. Then, 
clearly, we have 
\be 
\langle N^2(t) \rangle \; = \; \sum_{x_1} \sum_{x_2} \, \ob{G}^{(3)}( x_1,x_2,t) . 
\label{eq:aver_N2_01} 
\ee 
For $t > 0$, the three-point correlation function $\ob{G}^{(3)}$ satisfies the 
equation 
\be 
\ob{G}^{(3)}(x_1,x_2,t) = (1- f_t)^2 \sum_{e_1, e_2 = \pm 1} \, 
\frac{1}{4}\, \ob{G}^{(3)}(x_1 + e_1,x_2 + e_2,t-1) 
\label{eq:aver_G3_01} 
\ee 
for  $x_1 \neq x_2$.  We have to supplement 
this equation with the boundary condition for $x_1 = x_2$: 
\be 
\ob{G}^{(3)}( x_1,x_1,t) = \ob{G}(x_1,t) . 
\label{eq:aver_G3_02} 
\ee 
As for the case with no disorder, Eq.(\ref{eq:aver_G3_01}) 
is solved by the ansatz 
\be 
\ob{G}^{(3)}(x_1,x_2,t) = \sum_{u=0}^t \sum_{y} \, f(y,u) \, 
\ob{G}(x_1,t|y,u) 
\, 
\ob{G}(x_2,t|y,u) 
\label{eq:aver_G3_03} 
\ee 
which satisfies Eq.(\ref{eq:aver_G3_01}) automatically for $x_1 \neq x_2$.  
We choose the function $f(y,u)$ so that the equation holds also if $x_1 = x_2$.  
Putting $x_1 = x_2 = x$ in the previous equation, we get 
\be 
\ob{G}(x,t) = \sum_{u=0}^t \sum_{y} \, f(y,u) \, \ob{G}^2(x,t|y,u) 
\label{eq:aver_G_xt} 
\ee 
This is a linear equation connecting $f(y,u)$ to its values  
at sites higher up. Hence these can be solved recursively. 
The calculation  of $\langle N^2(t) \rangle$ can be further simplified.  
We define the function  
\be 
F(u) = \sum_y f(y,u), 
\ee 
and note that 
\be 
\sum_x \ob{G}(x,t|y,u) =  {\frac{1}{2}} \,  
{\cal{P}}_{t}\, / \, {\cal{P}}_{u-1} , 
\ee 
where we define ${\cal{P}}_{-1} =1$. 
Then using Eqs.(\ref{eq:aver_N2_01},\ref{eq:aver_G3_03}), we get  
\be 
\langle N^2(t)\rangle = \frac{1}{4} \,  
{\cal P}_t^2 \, \sum_{u=0}^t \; {\cal{P}}_{u-1}^{-2} \, F(u). 
\label{eq:aver_N2_02} 
\ee 
The recursion equation for $F(u)$ is also easily 
deduced from the recursion equation for the function $f(y,u)$. We note 
that  
$\sum_{x} \ob{G}^2(x,t|y,u) $  
is independent of $y$ and only depends on $t$ and $u$. Let us write 
\be 
\sum_{x} \ob{G}^2(x,t|y,u) = K(t,u). 
\ee 
Then summing Eq.(\ref{eq:aver_G_xt}) over $x$, we get 
\be 
{\frac{1}{2}} \, {\cal{P}}_{t} = \sum_{u=0}^t  \, F(u) \, K(t,u). 
\ee 
The effect of the disorder is just to introduce an overall rescaling 
factor to $K(t,u)$ in the absence of disorder: 
\be 
K(t,u) =  {\cal{P}}_{t}^2 \, {\cal{P}}_{u-1}^{-2} \; \sum_x \, G_0^2(x,t|y,u). 
\ee 
This implies that the asymptotic behavior of $F$ is given by  
\be 
F(u) \sim u ^{ - c - 1/2} . 
\ee 
Substituting in Eq.(\ref{eq:aver_N2_02}), we see that 
\be 
\langle N^2(t) \rangle \; \sim \; t^{1/2- c} . 
\ee 
The obtained scaling behavior of $\langle N(t) \rangle$  
and $\langle N^2(t) \rangle$ implies that the exponents of the  
assumed scaling form 
[Eq.(\ref{eq:scal_assump_01})] are given by   
$a = 1 + c$, and $b = 1/2$. 
This justifies the earlier heuristic calculation of the exponents.

It is straight-forward to extend this treatment to higher dimensions. 
We only mention the result here: in two (i.e. $2+1$ dimensions) and  
higher dimensions the avalanche exponents are given by 
\be 
\tau_t \; = \; 2 \, + \, c,  
\quad \quad 
{\rm and} \quad \quad 
\tau_m \; = \; (3 + c)/2 . 
\label{eq:exponent_higher_dim} 
\ee 
As in the case with no disorder, there are logarithmic 
correction terms to the exponent values given above 
in the two-dimensional case.

\section{Discussion} 
\label{sec:concluding_remarks} 
 
Let us comment briefly on the relationship of this model to that studied 
earlier by TD.  Their stochastic variant of the directed sandpile model is 
defined on the same lattice as presented in Fig.~\ref{fig:geo_01}, and 
particles are added on the top layer at random, and removed from below. 
The difference from the model studied here is that there is no quenched 
randomness, and the stochastic toppling rules are defined as follows:  If 
a particle is added to a site, only then is the site examined for 
toppling. If the number of particles exceeds $2$, with a non-$1$ 
probability $p$ a toppling occurs and one particle is transferred to each 
of the two downward sites, and with probability $(1-p)$ nothing happens. 
While there is particle conservation in the TD model, so far as the 
evolution of a single avalanche is considered, the sites where no toppling 
occurs when particles are added, act as sink sites. The density of sink 
sites is higher near the top in both models.  The randomness in the TD 
model provided by the stochastic toppling rules is `annealed', unlike the 
quenched randomness in the model studied here. 
 
In \cite{tadic}, it was shown that this model has a steady state for  
$p \geq p^{\star}$, where $p^{\star}$ is the critical probability for 
directed site percolation on a square lattice. In the steady state, there 
is a small fraction of sites having number of particles $2,3,4, \ldots$, but 
the average density is finite, and is a function of the depth of 
the layer.

TD argued  that a density less than the bulk density locally corresponds 
to a finite correlation length which is spatially varying. They argued if 
the medium has an inverse correlation length $\kappa(t)$, which is small, 
and varies slowly as a function of $t$, then the 
survival probability of a disturbance up to depth $t$ is given by  
\be 
{\rm Prob}( t) \; \sim \; {\rm Prob}_0 (t) \; \; 
\exp{\left ( -  \int_0^{t}  \kappa(t') dt' \right  ) }.  
\label{eq:ansatz} 
\ee  
The first factor is what we should get if 
$\kappa(t)$ is zero everywhere. The second term is a simple 
generalization of the $ \exp{(-\kappa t)}$ factor when $\kappa(t)$  
varies with $t$.  We note that the expression is non-perturbative, in that 
the small parameter $\kappa$ is exponentiated. More importantly, $\kappa$ 
is usually a nontrivial power of the starting perturbation $\delta 
\rho(t)$ in the density (in the TD case, $\kappa(t) \sim [\delta 
\rho(t)]^{\nu}$, where $\nu$ is the longitudinal correlation length of 
directed percolation problem).   
 
In the TD model,  $\kappa(t')$ varies as $c/t'$, and hence 
the argument of the exponential varies as $- c \, \log{t}$, for large $t$. 
This immediately gives the $c$-dependence of the power-law behavior  
of ${\rm 
Prob}(t)$. 
 
Using explicit calculation of correlation functions for small finite $t$, 
it is easy to verify that Eq.~(\ref{eq:ansatz}) is not exactly valid for 
the TD case. However, for the subsequent derivation the exponents of the 
TD model to be valid, the ansatz has to be at least asymptotically exact. 
As a proper treatment of the correlations in the steady state of TD model 
seems very difficult, one may try to find other simpler models where the 
validity of the ansatz can be demonstrated. 
 
The steady state in our model is very simple, and thus the model is a 
good testing-ground. It is gratifying that, in the present model, the 
ansatz does become exact, and the two-point correlation function satisfies 
Eq.~(\ref{eq:G_disorder}) exactly.  Thus we have shown that this 
approximation is exact for our model. 
 
The approximation is in the spirit of the familiar semiclassical 
Wentzel-Kra{\-}mers-Brillouin-Jeffreys approximation 
in quantum mechanics~\cite{WKBJ}.  We note that the  mathematical  
mechanism  giving rise to continuous variation of critical exponents with 
amplitude of perturbation applies to a much wider class of problems. 
Consider, for example, the generalization of the equilibrium 
problem treated  in \cite{hilhorst} 
involving extended  
surface defects to general $p$-vector spins $\{{\bf 
S}({\bf n})\}$  
in $d$-dimensional half-space. Lattice sites are labelled 
by a $d$-dimensional integer vector ${\bf n} = (n_1,n_2,\ldots,n_d)$ with 
$n_1 >0$. 
The Hamiltonian of the system is  
\be 
{\mathcal H} = \sum  J({\bf n,n'}) \, {\bf S}({\bf n})\cdot {\bf S}({\bf n'}) 
\ee 
where the summation is over all nearest neighbor pairs of 
sites~${\bf n, n'}$. 
The coupling  constant $J({\bf n,n'})$ is assumed to depend 
only on $n_1$ (we may assume that $n_1 \leq n_1'$) as 
\be 
J({\bf n,n'}) = J^{\star} - C \, n_1 ^ {-\chi} 
\ee 
where $J^{\star}$ is the  value of coupling in the bulk, and $C > 0$. 
  
As mentioned earlier, the  simple renormalization argument due to 
Burkhardt \cite{burkh82}  shows that for $\chi > 1/\nu$,  
where $\nu$ is the (bulk)  correlation length exponent, this perturbation 
is irrelevant, and for $\chi < 1/\nu$, it is relevant. 
In the marginal case $\chi = 1/\nu$, we get the inverse correlation 
length  $\kappa(n_1) \sim C^{\nu}/n_1$. 
Let us consider the spin-spin correlation  
function $\langle {\bf S}({\bf n})\cdot {\bf S}({\bf n'}) \rangle $.  
For simplicity, we may take ${\bf n}$ 
and ${\bf n'}$ to differ only in the first coordinate. Then, it is easy  
to write an expression  for this correlation function using the ansatz 
Eq.~(\ref{eq:ansatz}).  
The mathematical problem then is the same as discussed in this 
paper, and one finds that in the marginal case $ \chi = 1/\nu$, and the 
critical exponent corresponding to the decay 
of this correlation function varies linearly with $C^{\nu}$.

Of course, the dramatic effect of the boundary-perturbations on the 
exponents of the avalanche size distribution is due to the fact 
that all avalanches are initiated at the top boundary in our model. 
If we add particles at a fixed  distance $t_0$ from the boundary, it is 
easy to see that large avalanches of duration much greater than $t_0$ 
have the same asymptotic behavior as the case $t_0 = 0$. However, 
shorter avalanches  find significantly fewer defects than if they 
had been added at the top. If we added particles 
everywhere in the bulk, the avalanche 
size distribution is averaged over different values of $t_0$, and in the 
thermodynamical limit is the same as  the case without defects. 
 
 
We thank M. Barma for a critical reading of the manuscript.  
S.~L\"ubeck would like to thank V.~B.~Priezzhev and 
A.~Hucht for useful discussions.



\end{document}